\documentclass[aps,prb,twocolumn,groupedaddress,showpacs,showkeys]{revtex4}
\usepackage{graphicx}
\usepackage{dcolumn}
\usepackage{bm}
\usepackage{amsmath}
\begin{document}

\title{\bf Magnetism without magnetic impurities in oxides ZrO$_2$ and TiO$_2$}

\author{Franti\v{s}ek ~M\'aca$^{1}$, Josef ~Kudrnovsk\'y$^{1}$,
V\'aclav ~Drchal$^{1}$, and  Georges ~Bouzerar$^{2,3}$}

\affiliation{
$^{1}$Institute of Physics ASCR, Na Slovance 2, CZ-182 21 Prague 8, Czech Republic\\
$^{2}$Institut N\'eel, 25 avenue des Martyrs, CNRS, B.P. 166 38042
Grenoble Cedex 09
France.\\
$^{3}$Institut Laue Langevin, B.P. 156 38042 Grenoble France.\\
}
\date{\today}

\begin{abstract}

\parbox{14cm}{\rm}

\medskip

We perform a theoretical study of the magnetism induced in
transition metal dioxides ZrO$_2$ and TiO$_2$ by substitution of
the cation by a vacancy or an impurity from the groups 1A or 2A of
the periodic table, where the impurity is either K or Ca. In the
present study both supercell and embedded cluster methods are
used. It is demonstrated that the vacancy and the K-impurity leads
to a robust induced magnetic moment on the surrounding O-atoms for
both the cubic ZrO$_2$ and rutile TiO$_2$ host crystals. On the
other hand it is shown that Ca-impurity leads to a non magnetic
state. The native O-vacancy does not induce a magnetic moment in
the host dioxide crystal.
\end{abstract}


\maketitle

\section{Introduction}

We are all used when we refer to magnetic materials to have in
mind systems which at least contain the essential brick, the
magnetic impurities such as Fe, Ni or Mn. However, it has been
reported several times that magnetism can also be observed in
materials which does not contain such magnetic impurities.
Ferromagnetism in this new class of materials is referred to as
"$d^{0}$" or intrinsic ferromagnetism. Experimentally, it has
been shown that thin films of HfO$_2$ and ZrO$_2$
\cite{CoeyNature,Coey2} exhibit very high Curie temperature
(above room temperature). Unexpected ferromagnetism has been also
reported in irradiated graphite \cite{Graphite1,Graphite2} and
even in La-doped hexaborides such as CaB$_6$ \cite{Boride}.
Theoretically, ferromagnetism was predicted in systems, where the
Fermi level lies in a flat band \cite{Tasaki,Mielke}.
Recently it has been shown that CaAs is a good candidate for flat
band ferromagnetism \cite{CaAs}.

However, the intrinsic ferromagnetism is often a subject of controversy.
For example, it has been argued that the ferromagnetism in CaB$_6$ is
due to the presence of magnetic impurities such as Fe \cite{Matsubayashi}.
In addition, intrinsic ferromagnetism often appear difficult to be stabilized
and controlled as it strongly depends on the sample history: materials
grown under similar conditions may lead to non-magnetic samples. Until now,
there is no clear understanding of the underlying mechanism which
is at the origin of the observed magnetism. From a theoretical
point of view, it has been shown that point defects as cation
vacancies may be at the origin of the magnetism in these
materials \cite{Stoneham,Sawatzky,Sanvito,Zunger}.
The vacancy induces in these materials local magnetic moments on the
neighbouring oxygen atoms which then interact with relatively extended
exchange couplings.
The oxygen vacancies were originally also suggested as a source of the
magnetism in HfO$_2$ \cite{CoeyHfO2}, but the density functional
calculations \cite{Sanvito} found no evidence for the formation of
magnetic moments around such defects.

The long-range ferromagnetic order is possible only if the exchange
couplings are extended enough (beyond the percolation threshold).
Osorio-Guillen {\it et al.} \cite{Zunger} have shown that in CaO
oxide the ferromagnetism induced by cation vacancies is possible only
if the density of vacancies is sufficiently large (above 5\%).
They found that the formation energy of a vacancy is relatively high, which
implies that the cation vacancy induced ferromagnetism is unlikely.

A progress has been done recently in the understanding of the "$d^{0}$"
ferromagnetism \cite{d0prl} in the framework of a minimal model which
includes on equal footing the effects of disorder and electron-electron
correlation effects.
The effect of the cation vacancy or non-magnetic impurity on neighbouring
oxygen orbitals (the correlated disorder) is the relevant part of the
theory. This study has predicted that very high Curie temperatures can
be reached for realistic parameters.
It was shown that the optimal situation for ferromagnetism
depends mainly on two parameters (i) the position of the impurity band
which should be on the edge to split from the valence band for the hole
doped case, and (ii) the density of holes per defect should be of the order
of three. In order to circumvent the difficulties of controlling intrinsic
defects such as vacancies, an alternative way which consists in the
substitution of the four-valent cation A$^{4+}$ in dioxides such as
AO$_2$ (A=Ti, Zr, Hf) by a single-valent cation of the group 1A of
the periodic table has been proposed.
The substitution of the four-valent host $A^{4+}$ by a single-valent
impurity $X^{+}$, where X$=$ K, Li, Na, or Cs, introduces three holes
per impurity.
However, we should emphasize that it is not granted that the system will
become magnetic and that the impurity band will be located near the
top of the valence band.

In this paper we shall address a question concerning the formation
of magnetic moments in the host dioxides doped by non-magnetic impurities.
To this end we perform an {\it ab initio} study of the effect of substitution
of the cation by an element of the group 1A or 2A.
We shall employ two different approaches, both based on the local density
approximation (LDA) to the density functional approach:
(i) the supercell approach using the full-potential linearised augmented
plane-wave method (FLAPW) \cite{flapw}, and (ii) the embedded cluster method
(ECM) \cite{ecm} based on the Green function approach as implemented recently
in the framework of the tight-binding linear muffin-tin orbital (TB-LMTO)
method \cite{book}.
The single impurity in the supercell with increasing sizes corresponds to
decreasing but still finite impurity concentrations.
On the other hand, the embedded cluster approach allows us to address the
problem of moment formation for a single impurity in the infinite
host crystal.
Both approaches, however, allow for electronic relaxations on the sites
neighboring impurity which is an essential requirement for the formation of
induced non-local magnetic moments such as the A-vacancy in AO and AO$_2$
oxides \cite{Sawatzky, Zunger, Sanvito}.
We have also investigated the case of two nearest-neighbor impurities
that have a common nearest neighbor oxygen site and the effect of group
2A impurities (such as Ca) which introduce only two holes per Ca-impurity
when substituting the A-cation in AO$_2$ oxides.

\section{Formalism}

Below we briefly summarize the supercell and ECM approaches used in the
present paper.

\subsection{The supercell approach}

We consider a tetragonal supercell consisting of two conventional
cells for the fluorite structure of ZrO$_2$ and four tetragonal
units for the TiO$_2$ rutile structure. It is constructed  by
doubling the area of the basis in the $xy$ plane, and in the case
of TiO$_2$ also along the $c$ axis. The supercell contains 8
molecular units of the host dioxide and its lattice parameter are
($\sqrt{2} a$, $\sqrt{2} a$, $a$) for the fluorite structure and
($\sqrt{2} a$, $\sqrt{2} a$, 2$c$) for the rutile structure. The
substitutional impurity is introduced in the centre of the
supercell XA$_{7}$O$_{16}$ which corresponds to an effective
impurity concentration of 12.5\%. We neglect the change of the
lattice constant due to impurity as well as local relaxations. We
have used the experimental lattice constants $a$=5.256~\AA~for
fluorite ZrO$_2$ \cite{zro2_ex} and $a$ = $b$ = 4.59~\AA, $c$ =
2.96~\AA~for the rutile TiO$_2$ \cite{tio2_ex}. We have also
considered larger supercells corresponding to lower effective
concentrations, namely, the XA$_{31}$O$_{64}$ supercell
(concentration of 3.125\%) with lattice parameters (2$a$, 2$a$,
2$a$) for the fluorite and (2$a$, 2$a$, 4$c$) for the rutile
structures, respectively.

The spin-polarized electronic structure was obtained by the FLAPW
method as implemented in the WIEN2k program package \cite{flapw}
with the Perdew-Wang exchange-correlation potential \cite{pw92}.
We employed a plane-wave cutoff energy $E_{cut}$ = 16 Ry and
40 special points in the irreducible wedge of the Brillouin zone of
the large supercell with 96 atoms and correspondingly higher number
of points for smaller supercells.

The local magnetic moments are obtained by intergrating over the
muffin-tin spheres.
Radii 2.35~a.u. and 1.7~a.u. for K/Zr and O in ZrO$_2$, respectively
and 2.0~a.u. and 1.6~a.u. for K/Ti and O in TiO$_2$, respectively
were used.

\subsection{The embedded cluster approach}

The embedded cluster method (ECM) is based on the Green function approach.
The solid is divided into a cluster containing several tens or hundreds
of atoms and the rest of the system.
The cluster consists of the central (impurity) atom (or central atoms, e.g.,
pair of impurities) and their neighbors whose potentials are allowed
to differ from the host atoms due to the presence of perturbation.
It is useful to make calculations for a sequence of gradually growing clusters
in order to check how calculated quantities converge to a well-defined value.

We first find the self-consistent solution for an ideal translationally
invariant host as described by the LDA-Hamiltonian $H^{(0)}$ and
corresponding Green function $G^{(0)}$.
The perturbation $V_{PP}$ is localized, limited to the cluster region
characterized by the projection operator $P$ on the states of the cluster.
 For the selfconsistent LDA-solution of the perturbed problem we need
only the PP-block of its Green function $G$ described by the Hamiltonian
$H=H^{0}+V_{PP}$, namely,
\begin{equation}
G_{PP}=[ (G^{(0)})^{-1} - V_{PP} ]^{-1}_{PP}
\end{equation}
which is evaluated in real space from the known values of $G^{(0)}_{PP}$
calculated directly in the reciprocal space.
As the atomic charges in the cluster region are different from those in
the host, we have to include the necessary corrections to the host Madelung
potentials.
We have implemented the embedded cluster method within the TB-LMTO scheme
\cite{book}.

\section{Results and discussions}

We have considered two oxides, ZrO$_2$ in its cubic (fluorite) phase
and TiO$_2$ in the rutile structure.
Two kinds of impurities were considered, namely vacancies and K-dopants
in both cases.
Test calculations were also done for Ca-dopants in TiO$_2$, and for the
nearest neighbor X$-$O$-$X complex in ZrO$_2$ (X=vacancy, K-impurity).

The results are summarized in Tables~\ref{t2} and \ref{t3} for
ZrO$_2$ and for TiO$_2$, respectively, using the FLAPW method for
two effective impurity concentrations of 3.125\%  and 12.5\%
corresponding to 96- and 24-atom supercells. In Table~\ref{t4} we
present the results for a single impurity in the cubic ZrO$_2$ as
calculated by the ECM approach with the charge selfconsistency for
59 sites of the embedded cluster (see Table~\ref{t1}).
\begin{table}[h]
\caption{Coordination spheres around the impurity X replacing Zr
atom located at the origin in ZrO$_2$. The symbols E denote empty
spheres used in the TB-LMTO calculations. We show the coordinates
$\vec{r}$ of only one of the atoms in the coordination sphere as
the other follow from the O$_h$ symmetry of the cluster
(multiplicity n).}
  \begin{tabular}{c|cccccc}
    \hline
 site          & X & O & E & Zr &  O & E \\
   \hline
 n & 1 & 8 & 6 & 12 & 24 & 8 \\
    \hline
 $\vec{r}$ & ~$   (000)        $~ & ~$\frac{a}{2}(111)$~ & ~$\frac{a}{2}(200)$~ &
              ~$\frac{a}{2}(220)$~ & ~$\frac{a}{2}(311)$~ & ~$\frac{a}{2}(222)$~ \\
    \hline
 \end{tabular}
 \label{t1}
\end{table}

In both approaches we have found no induced magnetic moment
in the host due to native O-vacancies either in ZrO$_2$ or in TiO$_2$.
This is in agreement with results of Das Pemmaraju and Sanvito \cite{Sanvito}
for O-vacancies in HfO$_2$.

\subsection{Impurities in ZrO$_2$}

As seen from Table~\ref{t2}, both vacancy and K-impurity on
Zr-cation sublattice give rise to a well defined robust magnetic
moment induced on neighboring shells of O-atoms. Note that the
value of this moment could vary with the impurity concentration.
We first discuss the case of a vacancy. For relatively low
concentration of 3.125\% the total induced moment in the supercell
is 4~$\mu_{B}$. We found that the induced moment decreases only
slightly to 3.86~$\mu_{B}$ for a higher concentration of 12.5\%.
The relevant part of the induced magnetic moment is mainly located
on the first shell of host O-atoms. The induced moments on the
next oxygen shell are an order of magnitude smaller. The moment on
Zr-atoms neighbouring the defect are negligible. Despite the
smallness of induced moments on O-sites, the  total moment is
large because of the number of equivalent atoms in each shell
neighboring the impurity is large (8 in the first shell, 24 in the
second shell). These results are in a good agreement with those of
Ref.~\onlinecite{Sanvito} for vacancies in the monoclinic HfO$_2$
crystal. Of course, the magnitude of the induced moments could be
quantitatively influenced by possible lattice relaxation as
discussed in Ref.~\onlinecite{Sanvito}.

\begin{table}[h]
\caption{The total magnetic moment per cell (M$_{\rm cell}$) and
local magnetic moments on the impurity (m$_{\rm X}$) and on the
nearest-neighbor host atoms adjoining it for two cubic supercells,
XZr$_{7}$O$_{16}$ and XZr$_{31}$O$_{64}$, corresponding to
effective impurity concentrations of 12.5\% and 3.125\%,
respectively. Here m$_{\rm O1}$, m$_{\rm O2}$, and m$_{\rm Zr1}$
denote atoms in corresponding nearest-neighbor shells of the host.
The impurity atoms are X=vacancy and K. All magnetic moments are
given in units of $\mu_{B}$. }

  \begin{tabular}{c||c|c|c|c|c}
    \hline
impurity X& m$_{\rm X}$ & m$_{\rm O1}$ & m$_{\rm O2}$ & m$_{\rm Zr1}$ & M$_{\rm cell}$ \\
    \hline \hline
    vacancy (~12.5\%)& $-$  & $0.40$ & $0.05$ &  $-0.03$  & $3.81$ \\
    vacancy (3.125\%) & $-$  & $0.38$ & $0.03$ &  $-0.01$  & $4.00$ \\\hline
       K (~12.5\%)   & $0.10$  & $0.30$  & $0.05$  & $-0.02$  & $2.99$  \\
       K (3.125\%)   & $0.08$ & $0.21$ & $0.03$ & $-0.01$ & $2.67$ \\    \hline
  \end{tabular}
 \label{t2}
\end{table}

We now discuss the case of K-cation impurity. In this case we have
also found an induced moment but slightly smaller than that of
vacancy. Here, on the contrary, the system tends to a halfmetallic
behavior for sufficiently high concentration of defects. The total
induced moment slightly increases with the impurity concentration.
As expected the total induced moment is smaller roughly by one
$\mu_{B}$ smaller as compared to that of vacancy. Indeed,
K-impurity brings only three holes as compared to four holes in
the case of a vacancy. Similarly to the case of vacancy,  magnetic
moments induced on more distant O-atoms are an order of magnitude
smaller but they are nevertheless important as they help to
overcome the magnetic percolation threshold. The magnetic moments
on other sites (impurity and neighboring Zr atoms) are again
negligible. Because the induced moment around the cation
impurities is rather extended, the corresponding percolation
threshold is expected to be much smaller as compared to that of
conventional local magnetic moments e.g. in GaAs:Mn or GaN:Mn
alloys. This is a particularly interesting result for the
substitutional K-cation impurity which can be more easily
controlled experimentally than the intrinsic defects.

It is interesting to compare the results of supercell calculations
corresponding to small but finite impurity concentrations with
those obtained for a single impurity by the ECM approach (see
Table ~\ref{t4}). The total induced magnetic moment in the
insulator should be, strictly speaking, 4~$\mu_{B}$ and 3$\mu_{B}$
for vacancy and K-impurity, respectively. This is fulfilled very
well in the case of vacancy but not so accurately for K-impurity.
This indicates that a small part of the induced moment resides
outside the chosen cluster. We note that the results of the ECM
calculations could be also interpreted as the condition for the
formation of a local induced moment in the system (the
single-impurity limit).

In the framework of the supercell approach we have also investigated the
possibility of induced magnetic moments due to Ca-substitutional impurity
in ZrO$_2$.
We have found no induced moment in this case.
This result is in agreement with a general conclusion made in \cite{d0prl},
namely that the optimal situation for magnetism in dioxides requires
impurities which release three holes per defect into the valence/impurity band.
Thus, we find a decreasing tendency to form an induced magnetic moment in
ZrO$_2$ if Zr is substituted by vacancy, K, and Ca.
This result can be understood qualitatively as a consequence of decreasing
the charge difference between the host and impurity cation.
Such a qualitative explanation could be supported by a simple Stoner-picture model.
\begin{figure}[ht]
\includegraphics[width=6.5cm]{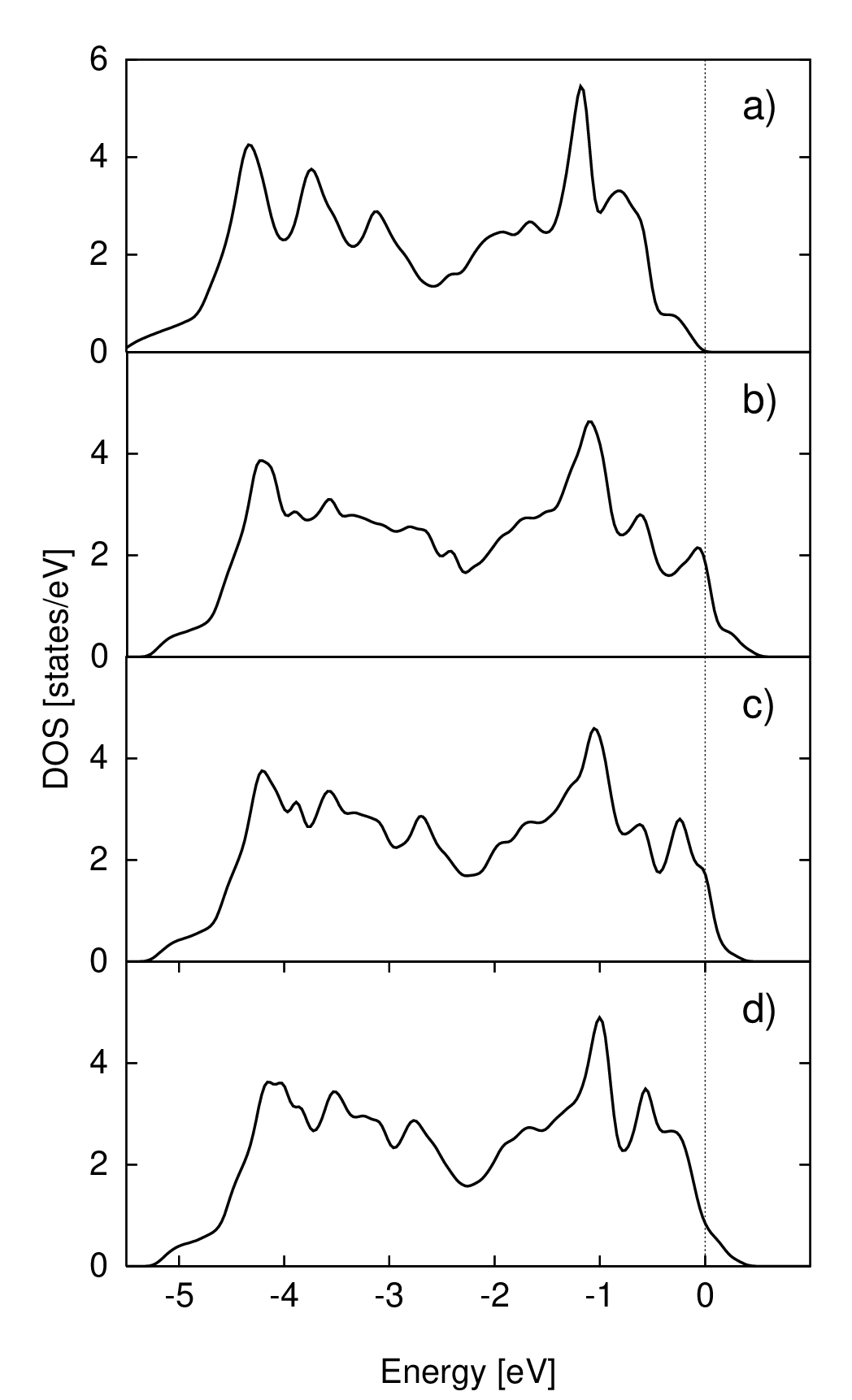}
\caption{The total densities of states per formula unit for
various impurities in the cubic nonmagnetic ZrO$_2$: (a) ZrO$_2$
host; (b) vacancy in ZrO$_2$; (c) K in ZrO$_2$; and (d) Ca in
ZrO$_2$. Calculations were performed for the supercell
XZr$_{15}$O$_{32}$ corresponding to the effective impurity
concentration of 6.25\%. The energy zero indicates the Fermi
energy.} \label{f1}
\end{figure}
We present in Fig.~\ref{f1} the total densities of states (DOS) of the
nonmagnetic phase calcualted using the supercell approach in the limit
corresponding to the effective impurity concentration of 6.25\%
(XZr$_{15}$O$_{32}$, X=vacancy, K, Ca).
A strong reduction of the total density of states at the Fermi energy for
Ca-impurity as compared to K-impurity and vacancy is clearly seen.
This explains the decreasing tendency to form an induced magnetic moment.

Using the ECM approach we have also investigated the case of two
nearest-neighbor vacancies/K-impurities in the ZrO$_2$ host that
have two equivalent nearest-neighbor oxygen sites. The total
induced moment on O-sites inside the cluster is strongly enhanced
when compared to the case of a single impurity: 7.1~$\mu_{B}$ for
two vacancies and 4.85~$\mu_{B}$ for two K atoms. This should be
compared with the values of 3.93~$\mu_{B}$ and 2.62~$\mu_{B}$
found for a single vacancy and K-impurity (see Table~\ref{t4}).
\begin{table}
\caption{The total magnetic moment inside the cluster M$_{\rm
clust}$ and local magnetic moments on the impurity (m$_{\rm X}$)
and on the nearest neighbor host atoms adjoining it for the cubic
XZrO$_{2}$ crystal with a single impurity calculated using the
embedded cluster method. Here m$_{\rm O1}$, m$_{\rm O2}$, and
m$_{\rm Zr1}$ denote atoms in corresponding nearest-neighbor
shells of the host. The impurity atoms are X=vacancy, and K. All
magnetic moments are given in units of $\mu_{B}$. }
  \begin{tabular}{cccccc}
    \hline
 impurity X & m$_{\rm X}$ & m$_{\rm O_{1}}$ & m$_{\rm O_{2}}$ & m$_{\rm Zr_{1}}$ & M$_{\rm clust}$ \\
    \hline
    vacancy & 0.005 & 0.421 & 0.027 & $-$0.007 & 3.927 \\
       K    & 0.158 & 0.244 & 0.023 & $-$0.005 & 2.616 \\
    \hline
  \end{tabular}
 \label{t4}
\end{table}
The local moment on O atoms neighboring the impurity is strongly enhanced,
namely 1.22~$\mu_{B}$ and 0.77~$\mu_{B}$ for the vacancy and K-impurity,
respectively. The other induced moments are similar to those found for
a single impurity case. One could conclude that the nearest-neighbor
X-O-X cluster (X=vacancy or K) induces in ZrO$_2$ host a larger and more
extended magnetic moment.

\begin{figure}[tbp]
\includegraphics[width=6.5cm]{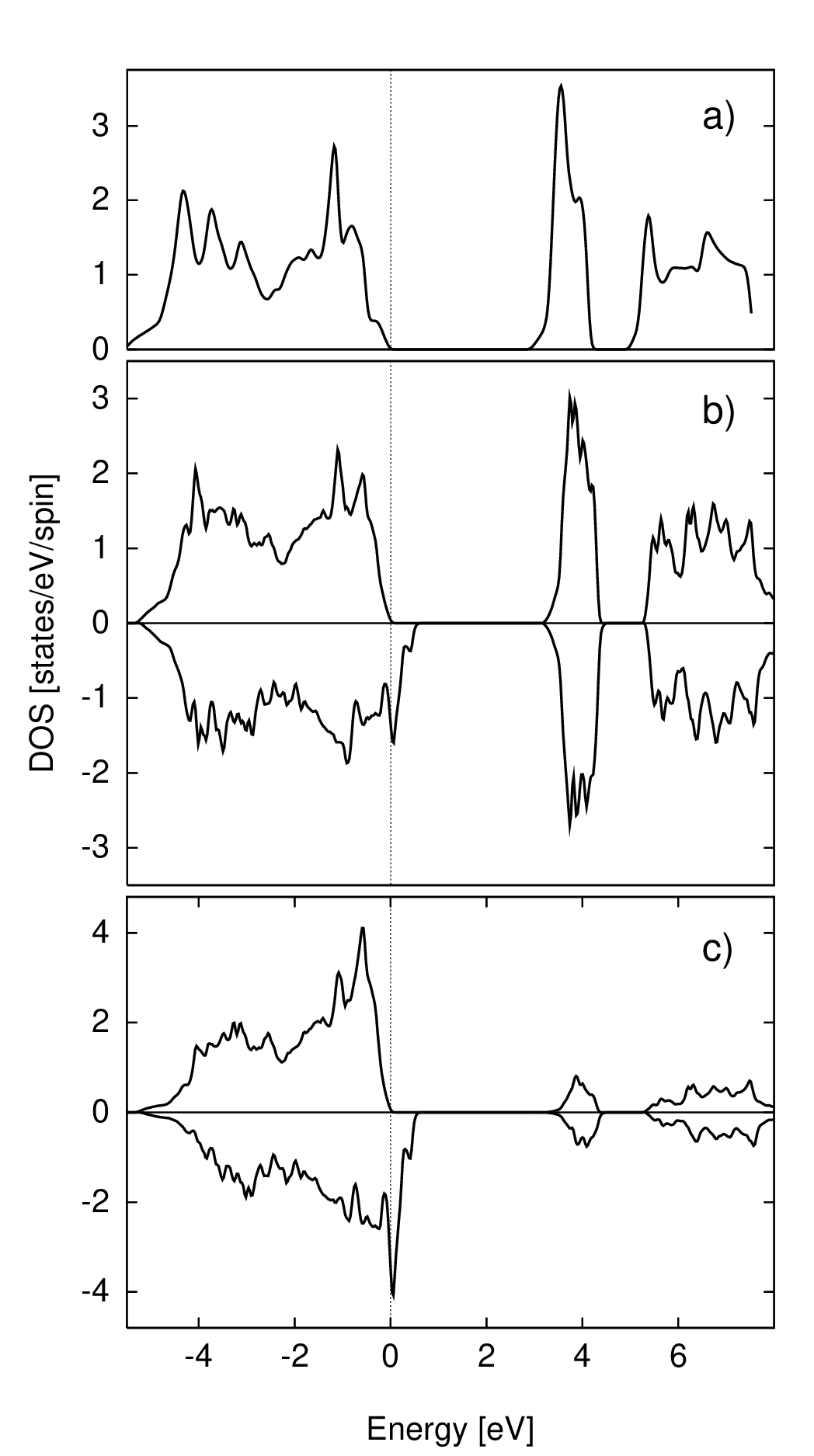}
\caption{(a) The total density of states per formula unit for
ZrO$_2$ host and (b) the spin resolved densities of states fro
KZr$_7$O$_{16}$ supercell. (c) The spin-resolved local densities
of states per formula unit on the first O-shell surrounding the
impurity  for the cubic KZr$_7$O$_{16}$ supercell. The effective
K-impurity concentration is 12.5\%. The energy zero indicates the
Fermi energy.  } \label{f2}
\end{figure}
The spin resolved total DOS and local DOS on O-sites for substitutional
K-impurity in ZrO$_2$ are shown in Fig.~\ref{f2} for
the KZr$_{7}$O$_{16}$ supercell (12.5\% of impurities).
The substitutional K-impurities in ZrO$_2$ polarize the
host band giving rise to well split spin-subbands.
The induced impurity peak close to the valence band edge is seen in
the minority DOS while the majority DOS is influenced by
impurities only negligibly (compare Fig.~\ref{f2}a and \ref{f2}b).
This effect is even more clearly seen in the O-local DOS shown in
Fig.~\ref{f2}c from which it is found that the valence band is
formed predominatly by the oxygen $p$-states.
The above results are again in a qualitative agreement with those of
Ref.~\onlinecite{d0prl}. Namely, the presence of an impurity band
close to the top of the valence band is a precursor for the appearence
of magnetism in dioxides. This is the case for both a vacancy and
substitutional K-impurity (see Fig.~\ref{f1}b,c).
In contrast, for Ca substitution, the impurity band is shifted deeper
into the valence band, which prevents the formation of a magnetic
moment.

\subsection{Impurities in TiO$_2$}

We have also investigated the possibility for non-magnetic
impurity induced magnetism in TiO$_2$. We used similar supercells,
namely XTi$_{7}$O$_{16}$ and XTi$_{31}$O$_{64}$ as in the case of
ZrO$_2$ (effective concentrations are 12.5\% and 3.125\%). The
main results (see Table~\ref{t3}) are in qualitative agreement
with those for the cubic ZrO$_2$ structure.  Nevertheless, for
higher concentration of vacancies (12.5\%) no magnetic state has
been found in TiO$_2$. For smaller concentrations both vacancy and
substitutional K-impurity induce magnetic moments on the
surrounding O-sites and the size of total induced moments are
similar in ZrO$_2$ and TiO$_2$. There are, however, some
quantitative differences due to different lattices. In the rutile
structure there are four equivalent nearest-neighbor O-sites and
two other next nearest-neighbor O-sites with almost the same
distance from the Ti-/impurity site. Lattice anisotropy leads to a
different distribution of the induced moments.
\begin{table}[h]
\caption{The total magnetic moment per cell (M$_{\rm cell}$) and
local magnetic moments on the impurity (m$_{\rm X}$) and on the
nearest neighbors host atoms adjoining it for two cubic
supercells, XTi$_{7}$O$_{16}$ and XTi$_{31}$O$_{64}$,
corresponding to effective impurity concentrations of 12.5\% and
3.125\%, respectively. Here m$_{\rm O1}$, m$_{\rm O2}$, and
m$_{\rm Ti_{1}}$ denote atoms in corresponding nearest-neighbor
shells of the host. The impurity atoms are X=vacancy, and K. All
magnetic moments are given in units of $\mu_{B}$. }

  \begin{tabular}{c||c|c|c|c|c}
    \hline
 impurity X & m$_{\rm X}$ & m$_{\rm O_{1}}$ & m$_{\rm O_{2}}$ & m$_{\rm Ti_{1}}$ & M$_{\rm cell}$ \\
    \hline \hline
    vacancy (~12.5\%)& $-$  & 0.00  & 0.00  & 0.00  & 0.00  \\
    vacancy (3.125\%)& $-$  & 0.44 & 0.28  & $-$0.04 & 3.46 \\\hline
       K (~12.5\%)   & 0.27  & 0.05  & 0.32  & 0.00  & 2.10  \\
       K (3.125\%)   & 0.35 & 0.31 & 0.18 & 0.01 & 2.91 \\
    \hline
  \end{tabular}
 \label{t3}
\end{table}

We have also checked the effect of local structure relaxations in
the vicinity of impurity for the case of KTi$_{7}$O$_{16}$. The
distance between the K-impurity and four nearest neighbor O1
oxygen sites increases from 1.95~\AA~ to 2.27~\AA~ and between two
O2 oxygen sites from 1.98~\AA~ to 2.15~\AA~. The existence of
induced moment due to K-impurity is not influenced, the total
induced moment changes from 2.10~$\mu _B$ to 2.08~$\mu _B$, the
local moments on O1-sites change from 0.32~$\mu _B$ to 0.26~$\mu
_B$ while the local moments on O2-sites remain unchanged
(0.05~$\mu _B$).

The induced local moments are different for different impurity
concentrations, e.g. on O2 sites is local magnetic moment
0.19~$\mu_{B}$ for a concentration of 3.125\% but becomes
$-$0.05~$\mu_{B}$ when the impurity concentration is 12.5\%. The
reason for this variation is not clear to us. For example, one
could speculate that this could be a result of different hole
concentrations in both spin-subbands. We should also mention a
relatively large local moment on K-impurity which appear to be
comparable to that induced on individual O-sites. It is few times
larger than the local K-moment in ZrO$_2$ (see Table~\ref{t2}).

\begin{figure}[tbp]
\includegraphics[width=6.5cm]{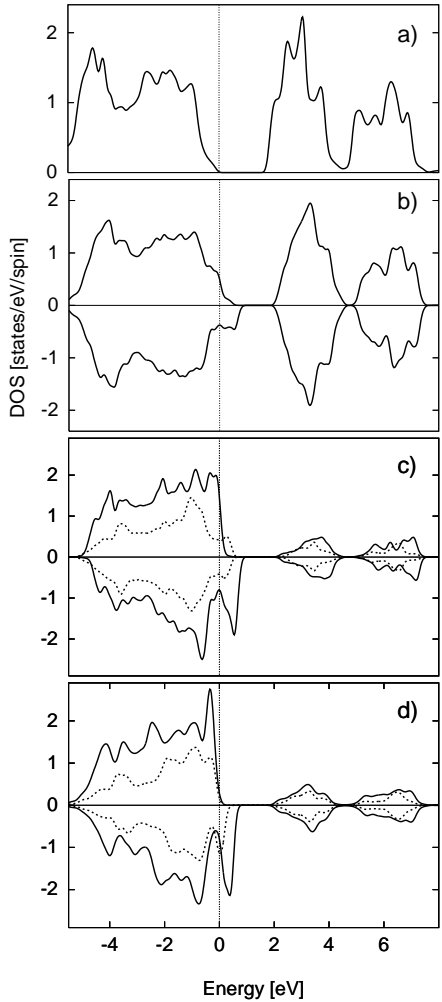}

\caption{The total densities of states per formula unit for (a)
TiO$_2$ host with the rutile structure and (b) spin resolved
density of states for KTi$_7$O$_{16}$. The local spin-resolved
densities of states on the first two O-shells surrounding the
impurity in the ferromagnetic cubic (c) KTi$_7$O$_{16}$ and (d)
KTi$_{15}$O$_{36}$ that correspond to concentrations of 12.5\% and
6.25\% ,respectively. The energy zero indicates the Fermi energy.}
\label{f3}
\end{figure}

The total and local DOS for KTi$_7$O$_{16}$ supercell are shown in
Fig.~\ref{f3}.
The general trend is similar to that found for K-impurities in ZrO$_2$:
there is an impurity band in the minority DOS but less pronounced
thus indicating a slightly weaker disorder due to K-impurities in
TiO$_2$ as compared to ZrO$_2$ (in terms of {\it model language} of
Ref.~\onlinecite{d0prl}).
The majority band is again only weakly perturbed and the system is metallic
with the Fermi energy lying inside the valence band for both the majority
and minority bands.

The spin resolved local DOS on O2 is weakly spin dependent for
KTi$_7$O$_{16}$ supercell (see Fig.~\ref{f3}c).
This results in a very small induced moment on these
sites ($-$0.05~$\mu_{B}$).
On the other hand, the O1 local DOS shows a halfmetallic behavior and
gives  rise to a larger induced moment (0.28~$\mu_{B}$).
For lower concentration we observe a halfmetallic behavior for both
O1 and O2 spin-resolved DOS (see Fig.~\ref{f3}d). This results in a similar
size for the moment on O1 atoms but a significantly increased O2 moment
(see Table~\ref{t3}).

\subsection{Local magnetization around impurities in ZrO$_2$}

\begin{figure}[htbp]
\includegraphics[width=7.5cm, angle=-90]{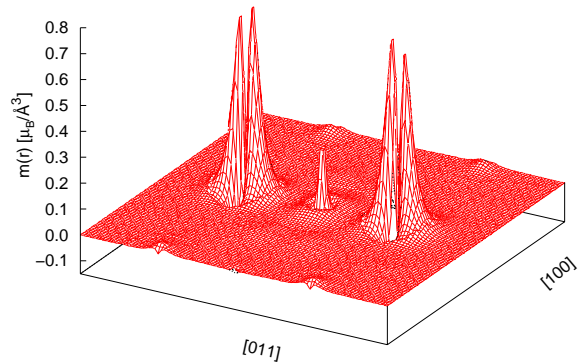}
\caption{The local magnetization for the ferromagnetic cubic
KZr$_7$O$_{16}$ supercell in the [110]-plane. The K-impurity is in
the center and two nearest neighbor O-atoms in the first O-shell,
while four O-atoms in the second O-shell are visible at [011]
edges.} \label{f4}
\end{figure}

The local magnetization $m({\bf r})$ in [110]-plane around
the impurity in real space is shown in Fig.~\ref{f4} for
KZr$_7$O$_{16}$ supercell.
The small local moment on K is in the centre of the plot with
strongly damped Friedel-like oscillations. Such a damping is due
to halfmetallic behavior of this system, i.e., the position of the
Fermi energy lies in the gap of the majority band. This feature is
similar to that observed in the magnetic exchange couplings in the
diluted magnetic semiconductors \cite{eiour}. Large increase of
the magnetization density in the neighborhood of the O-sites
indicates the existence of an induced moment, few times larger
than the local K-moment (see Table~\ref{t2}). The induced moments
on the second shell of O-sites are also shown. Their magnitude is
small but because the number of sites in the shell is relatively
high (24), their contribution is non-negligible.\\

\section{Conclusions}

In conclusion, we have shown successfully that a new path to
controlled $d^0$-ferromagnetism can be achieved by cation doping
of oxides with well defined elements of the periodic table.
Specifically, we have shown that the promissing systems are both
cubic ZrO$_2$ and rutile TiO$_2$ materials doped with K (group
1A). The methods which have been used are both the supercell
approach and the embedded cluster method. The main conclusions of
our study are: (i) The magnetic moment is mainly induced on the
first oxygen shell neighbouring the defect. (ii) A resonance peak
near the top of the valence band in the minority channel is found,
while the majority channel remains mainly unperturbed. This result
is in agreement with a recent theoretical study based on model
calculations \cite{d0prl}. (iii) The substitution by Ca (group 2A)
has lead to a non magnetic ground state. (iv) We have verified for
a specific case that the total induced moment is robust with
respect to lattice relaxations. (v) Our results also indicate that
these effects are not inherent to a specific lattice structure.

The interesting and crucial issues which have to be addressed in the future
are the impurity formation energies and the exchange integrals between
induced moments.

This work has been done within the project AVOZ1-010-0520 of the
AS CR. We acknowledge the support from the Grant Agency of the
Czech Republic, Contract No. 202/07/0456,  COST P19-OC150 project.
and support from the Grant Agency of the Academy Sciences of the
Czech Republic (A100100616).

\end{document}